\documentclass[cameraready]{Interspeech}

\title{RAS: a Reliability Oriented Metric for Automatic Speech Recognition}
\author[affiliation={1}, equalcontribution, orcid=0009-0001-9648-8842]{Wenbin}{Huang}
\author[affiliation={1}, equalcontribution, orcid=0009-0004-2806-2024]{Yuhang}{Qiu}
\author[affiliation={1}, orcid=0009-0004-4428-4065]{Bohan}{Li}
\author[affiliation={1}, orcid=0009-0003-8114-2085]{Yiwei}{Guo}
\author[affiliation={1}, orcid=0009-0002-0858-1527]{Jing}{Peng}
\author[affiliation={1}, orcid=0009-0008-7959-0336]{Hankun}{Wang}
\author[affiliation={1}, orcid=0000-0001-7423-617X]{Xie}{Chen}
\author[affiliation={1}, correspondingauthor, orcid=0000-0002-7102-9826]{Kai}{Yu}

\address{
    $^1$ X-LANCE Lab, School of Computer Science,  Shanghai Jiao Tong University, China \\
    $^1$ MoE Key Lab of Artificial Intelligence; Jiangsu Key Lab of Language Computing, China
}

\email{\{hartmann\_psi, qiuyuhang, kai.yu\}@sjtu.edu.cn}


\usepackage{comment}
\usepackage{xcolor}
\usepackage[most]{tcolorbox}
\usepackage{multirow}
\definecolor{skyblue}{RGB}{135,206,235}
\usepackage{float}

\begin{document}

\maketitle

\begin{abstract}


    Automatic speech recognition systems often produce confident yet incorrect transcriptions under noisy or ambiguous conditions, which can be misleading for both users and downstream applications. Standard evaluation based on Word Error Rate focuses solely on accuracy and fails to capture transcription reliability. We introduce an abstention-aware transcription framework that enables ASR models to explicitly abstain from uncertain segments. To evaluate reliability under abstention, we propose RAS, a reliability-oriented metric that balances transcription informativeness and error aversion, with its trade-off parameter calibrated by human preference. We then train an abstention-aware ASR model through supervised bootstrapping followed by reinforcement learning. Our experiments demonstrate substantial improvements in transcription reliability while maintaining competitive accuracy.
\end{abstract}

\section{Introduction}
Modern automatic speech recognition (ASR) systems achieve high accuracy in clean acoustic conditions, yet they often still produce superficially fluent transcripts in the presence of noise, overlapping speech, signal degradation, or low-resource settings. These outputs are frequently the result of forced decoding under weak acoustic evidence, yielding errors that appear confident rather than explicitly uncertain. Such \textit{plausible-but-wrong} transcriptions can mislead downstream decision-making. Under ambiguous acoustics, neither humans nor language models can always make reliable corrections. And, due to their readability, may further reduce human vigilance during review. The risk is especially acute in high-stakes applications with stringent transcription requirements, such as medical documentation and legal records. However, dominant evaluation paradigms do not adequately characterize this failure mode.\footnote{Code available at: \url{https://github.com/HartmannPsi/Reliability-Aware-Score}}

While abstention (i.e., learning with a reject option) and selective prediction have been shown to improve reliability in machine learning by explicitly trading coverage for reduced risk \cite{Optimal-Reject, Calibrated-Structured-Prediction, xu2024reducing, xu2024rejection, Schnwlder2025AbstentionIA, zheng2025enhancing}, these approaches typically operate at the instance level, rejecting entire utterances. This is unsuitable for ASR, where discarding a whole sentence eliminates valuable partial information. A related line of work, ASR confidence estimation \cite{Jiang2005ConfidenceMF, Onea2021AnEO, Futami2021ASRRA, Naowarat2023WordlevelCE, Huo2025IdentifyingAC} constitutes a viable avenue for addressing this issue. However, most existing methods still follow a two-stage, post-hoc paradigm: they produce transcriptions first and subsequently assign uncertainty scores as a separate metadata layer. Such implicit modeling lacks an internal mechanism for the model to actively ``opt-out'' of unreliable segments. Meanwhile, standard \textit{Word-Error-Rate}(WER) \cite{lawrence-overview, nist_hub5ne_1997} and its edit-distance variants\cite{edit-distance, K2024AdvocatingCER, MER_WIL}, as well as semantic metrics\cite{bleu, ROUGE, bertscore}, all implicitly assume forced transcription and do not explicitly model abstention or reliability under selective prediction. Consequently, these metrics cannot assess whether an ASR system makes an appropriate trade-off between informativeness and reliability under uncertainty.

\begin{figure}[!tbp]
    \centering
    \includegraphics[width=0.8\columnwidth]{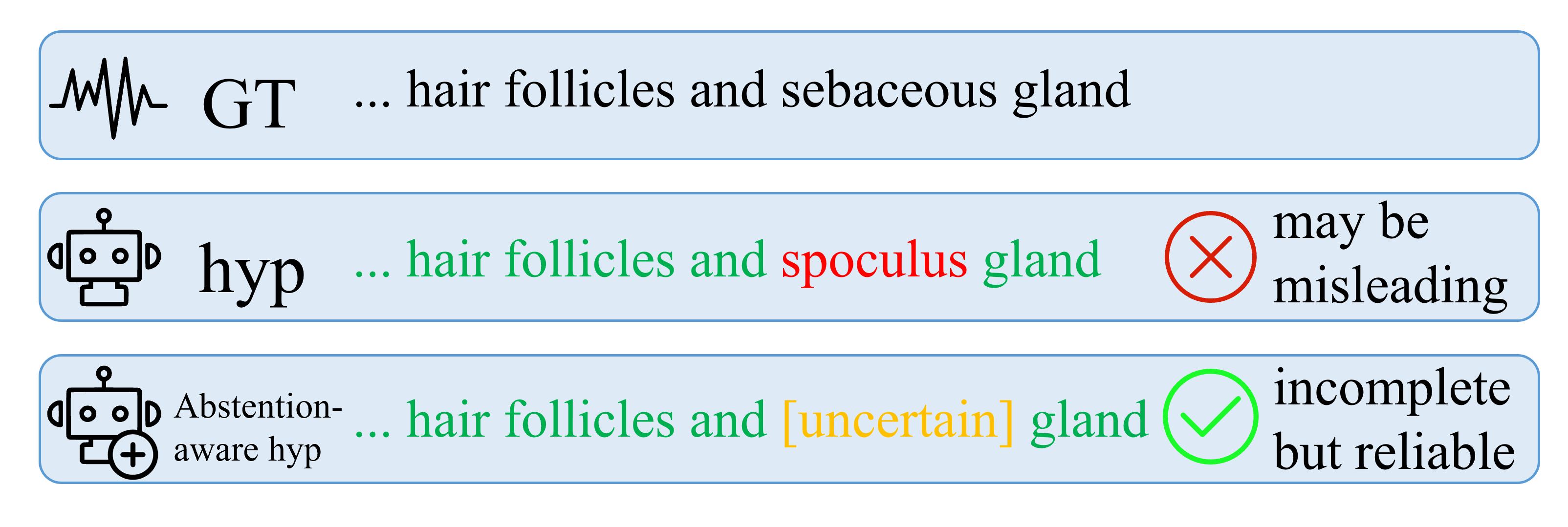}\vspace{-8pt}
    \caption{
    Conventional v.s. abstention-aware hypothesis.
    \vspace{-9pt}
    }
    \label{fig:reliability_comparison}
\end{figure}

In this work, we argue that reliable ASR should be able to to explicitly abstain from localized uncertain predictions. We introduce a fine-grained abstention paradigm that augments the output space with a dedicated placeholder, as shown in Figure \ref{fig:reliability_comparison}. Unlike full-sentence rejection, our approach allows the model to selectively abstain on ambiguous segments while transcribing the remaining content with high commitment. This transition from passive scoring to active, fine-grained abstention allows downstream modules to treat these placeholders as missing information rather than misleading hallucinations.
To evaluate this paradigm, we propose \textbf{R}eliability-\textbf{A}ware \textbf{S}core (RAS), a novel metric derived from a modified edit distance that accounts for placeholder marks. RAS explicitly balances transcription usefulness and error aversion through a trade-off hyperparameter calibrated via human listening tests. This provides a principled objective for both the evaluation and optimization of reliability-aware ASR.

We further demonstrate the practical utility of our framework by developing an abstention-aware Whisper \cite{whisper} model. By employing a two-stage training pipeline—comprising supervised bootstrapping followed by reinforcement learning (RL) \cite{sutton1998reinforcement} using RAS as the reward—we significantly enhance system reliability. Experimental results demonstrate that our approach substantially improves transcription trustworthiness, particularly in low-resource and noisy conditions, while maintaining competitive accuracy. Our main contributions are summarized as follows:
\begin{itemize}
\item We extend selective prediction to sequential ASR, enabling segment-level abstention instead of full-utterance rejection.
\item We propose RAS, a reliability-oriented metric calibrated with human preference data to quantify the trade-off between informativeness and error aversion.
\item We establish a robust training pipeline combining supervised learning and RL, which significantly elevates the reliability of ASR models in challenging acoustic environments.
\end{itemize}
\section{Reliability-Aware Score}
To equip ASR models with explicit rejection capability, we extend the original vocabulary with a special placeholder mark $\mathcal{PH}$. Unlike ordinary lexical words, the $\mathcal{PH}$ represents abstention: when the model encounters low-quality or ambiguous speech and cannot reliably determine the underlying content, it outputs $\mathcal{PH}$ to indicate uncertainty rather than producing a potentially misleading guess. With the introduction of $\mathcal{PH}$, we now modify the standard WER into RAS, an abstention-aware metric, calculated by a proposed dynamic programming formulation that extends the definition of standard edit distance \cite{edit-distance}.

\subsection{Dynamic programming formulation}\label{section:dp}
Let the reference text $\mathit{ref}$ of length $N>0$ contains no $\mathcal{PH}$s, while the hypothesis $\mathit{hyp}$ of length $M$ may contain $\mathcal{PH}$s. Let $r_i$ and $h_j$ denote the $i$-th and $j$-th words (0-indexing) from $\mathit{ref}$ and $\mathit{hyp}$, respectively.
Unlike standard edit distance, a single $\mathcal{PH}$ may align to zero or multiple consecutive reference words, reflecting abstention over an uncertain acoustic segment. Consecutive $\mathcal{PH}$s are merged to avoid redundancy.

When extending the definition of edit distance, it is inappropriate to treat errors introduced by $\mathcal{PH}$s in the same manner as ordinary word errors. Although failing to provide useful information, a $\mathcal{PH}$ does not assert an incorrect lexical item and therefore avoids introducing potentially misleading content. In this sense, $\mathcal{PH}$-related errors are comparatively less harmful and more reliability-preserving. To reflect this distinction, we assign a cost factor $\alpha \in (0,1)$ to all $\mathcal{PH}$-related operations.
We therefore define $g^\alpha_{i,j}$ as the minimum weighted edit distance from $\mathit{ref}[0{:}i)$ to $\mathit{hyp}[0{:}j)$ under an abstention-aware alignment scheme, with boundary conditions: 
\begin{equation}
g^\alpha_{0,0}=0,\quad
g^\alpha_{i,0}=i,\quad
g^\alpha_{0,j}=\sum_{t=0}^{j-1}
\begin{cases}
1,& h_t\ne\mathcal{PH}, \\
\alpha,& h_t=\mathcal{PH},
\end{cases} 
\end{equation}
and define the transition as the minimum of three potential costs:
\begin{equation}
    g_{i,j}^{\alpha} = \min \{ \text{Sub}, \text{Del}, \text{Ins} \},
\end{equation}
where
\begin{align}
    \text{Sub} &= \begin{cases} 
        g_{i-1,j-1}^{\alpha} + \mathbb{I}_{r_{i-1} \neq h_{j-1}}, & h_{j-1} \neq \mathcal{PH} \\ 
        \min\limits_{0\le k <i} \{g_{k,j-1}^{\alpha} + \alpha(i-k)\}, & h_{j-1} = \mathcal{PH} 
    \end{cases}~,
    \\
    \text{Del} &= g_{i-1,j}^{\alpha} + 1, \\
    \text{Ins} &= \begin{cases} 
        g_{i,j-1}^{\alpha} + 1, & h_{j-1} \neq \mathcal{PH} \\ 
        g_{i,j-1}^{\alpha} + \alpha, & h_{j-1} = \mathcal{PH} 
    \end{cases}~,
\end{align}
which corresponds to standard edit operations when $h_{j-1} \neq \mathcal{PH}$, but introduces a flexible alignment for $\mathcal{PH}$s. When $h_{j-1} = \mathcal{PH}$, the substitution term $\text{Sub}$ allows $\mathcal{PH}$ to align with an arbitrary-length segment $r_{[k:i)}$ in the reference, effectively ``absorbing'' $i-k$ deletions at a discounted cost of $\alpha$ per word. This corresponds to a many-to-one mapping where $\mathcal{PH}$ covers a contiguous span of uncertain content. When no $\mathcal{PH}$ appears in $\mathit{hyp}$, the formulation reduces exactly to standard edit distance. We further define the weighted edit distance from $\textit{ref}$ to $\textit{hyp}$ as $g^\alpha(\textit{ref},\textit{hyp}) := g^\alpha_{N,M}$, computed in $O(N^2M)$ time.
\begin{figure}[t]
    \centering
    \includegraphics[width=0.8\columnwidth]{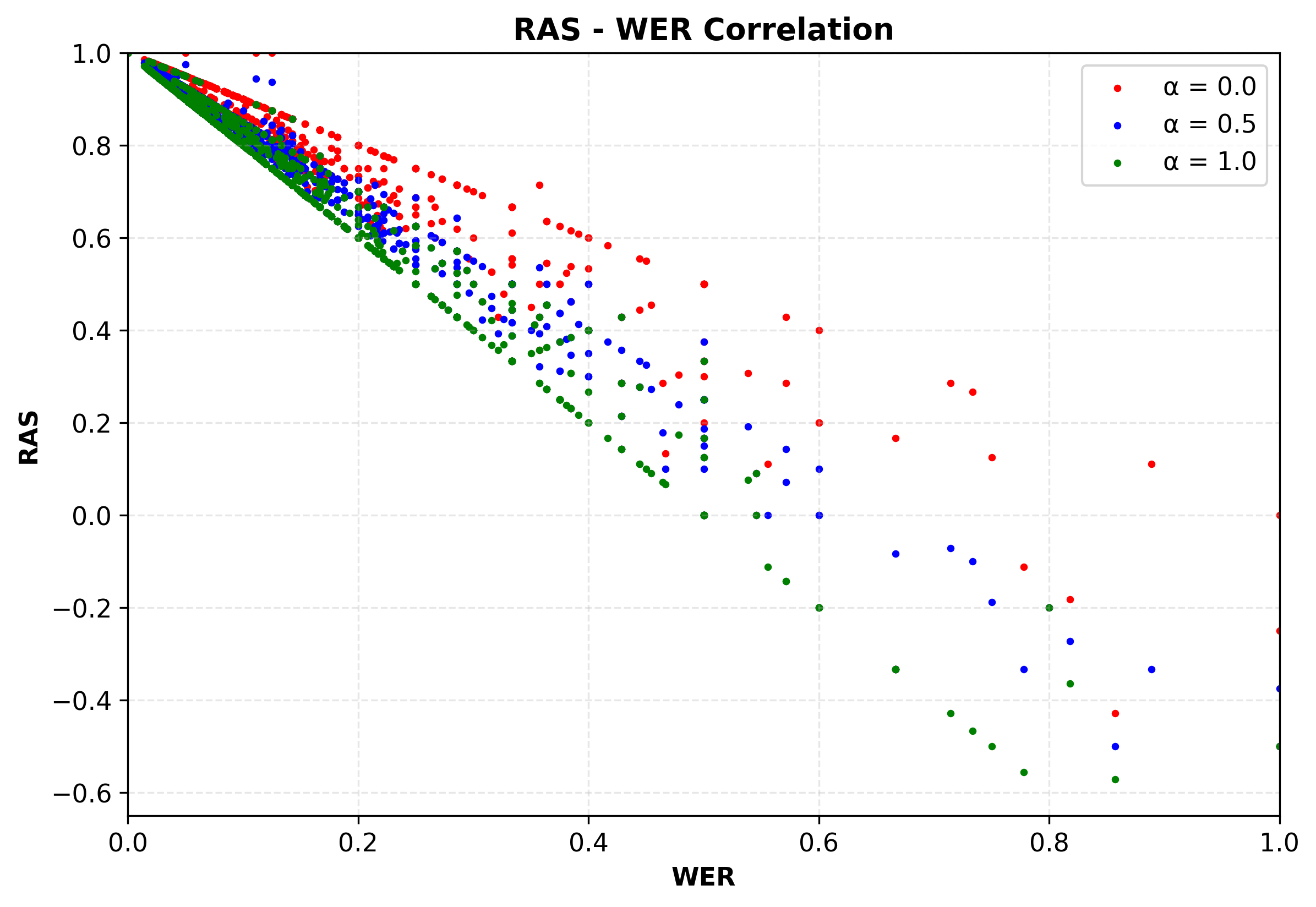}\vspace{-8pt}
    \caption{
    Relationship between WER and RAS under different $\alpha$ settings on LibriSpeech-test-clean \cite{librispeech} (Base+PH-Supv+RL) as in Section \ref{sec:results}. Each point corresponds to one utterance.\vspace{-7pt}
    }
    \label{fig:correlation}
\end{figure}
\subsection{Beyond WER: The definition of RAS}\label{sec:utility}
From the optimal DP alignment, we obtain the counts of correct matches $C(\textit{ref},\textit{hyp})$ and $g^\alpha(\textit{ref},\textit{hyp})$. Inspired by existing utility formulations for reliability-aware decision making in NLP \cite{xu-etal-2025-alignment}, we define RAS $R(\textit{ref},\textit{hyp})$ as:
\begin{equation}
R(\textit{ref},\textit{hyp}) = \text{Usefulness} - \text{Cost},
\end{equation}
where usefulness measures the proportion of correctly transcribed words, and cost penalizes alignment errors with differential weighting as described in Section \ref{section:dp}:
\begin{equation}
\text{Usefulness} = \frac{C(\textit{ref},\textit{hyp})}{N},
\quad\text{Cost} = \frac{g^\alpha(\textit{ref},\textit{hyp})}{N},
\end{equation}
with $\alpha \in (0,1)$. 
This formulation directly balances informative content against erroneous commitments under the proposed alignment. 
It has a maximum of 1, and a higher RAS implies a more reliable and useful ASR model.
When no $\mathcal{PH}$ appears in $\mathit{hyp}$, RAS reduces to be similar to standard WER:
\begin{align}
R(\textit{ref},\textit{hyp})&=1-\frac{2(S+D)+I}{N},\\
\text{WER}(\textit{ref}, \textit{hyp})&=\frac{S+D+I}{N},
\end{align}
both RAS and WER are monotonic functions of the total number of word errors $(S+D+I)$. Therefore, in the absence of $\mathcal{PH}$s, maximizing RAS is equivalent to minimizing WER in terms of optimization direction. The proposed objective holds a clear negative correlation with WER as shown in Figure \ref{fig:correlation}.

In practice, $\alpha$ should be calibrated to align with human preferences over reliability in specific application contexts, rather than selected heuristically.

\subsection{Human preference alignment via listening tests}\label{section:listening-test}
Listening tests were conducted using the BeaqleJS framework \cite{beaqlejs}. For each audio sample $i$, participants were presented with the ground-truth transcript ($G_i$) and two simulated ASR outputs: a conventional transcription without $\mathcal{PH}$s ($A_i$) and an abstention-aware transcription ($B_i$). Transcription $B_i$ was constructed by replacing recognition errors in $A_i$ with $\mathcal{PH}$s and additionally masking a small fraction of correct words to simulate conservative abstention. Participants were asked to select the transcription they considered more reliable, with an additional ``Can’t Decide'' option.

We estimate $\alpha$ from collected human preferences. Let the number of audios be $K$. For each audio $i$, let $k_i^A$, $k_i^B$, and $k_i^C$ denote the numbers of subjects preferring  conventional transcription $A_i$, abstention-aware transcription $B_i$, and indicating indifference, respectively, with $s_i = k_i^A + k_i^B + k_i^C$. Let $P_i$ denote the probability of human preferring $B_i$ for audio $i$, and $R$, $C$, $g^\alpha$, $N$ have the same meanings as in Section \ref{sec:utility}. We use notation $\Delta M_i:=M(G_i,B_i)-M(G_i,A_i)$ where $M\in\{R,C,g^\alpha\}$. Following the Bradley–Terry model \cite{bradley-terry}, $P_i$ is modeled as a function of $\Delta R_i$. Note that $R(G_i,A_i)$, $R(G_i,B_i)$ share the same denominator:
{
\begin{equation}
P_i=\sigma(\Delta R_i)=\sigma\left[\frac1{N_i}\left(\Delta C_i-\Delta g^\alpha_i\right)\right],
\end{equation}
}
where $\sigma(\cdot)$ denotes the logistic function. We define the preference likelihood loss and incorporate indifference responses as a regularization encouraging small RAS differences. The final objective is the weighted sum of the two:
\begin{gather}
\mathcal{L}_\text{pref}=-\frac1K\sum_i\left[\frac{k^B_i}{s_i}\log P_i+\frac{k^A_i}{s_i}\log(1-P_i)\right],\\
\mathcal{L}_\text{tie}=\frac1K\sum_i  \frac{k^C_i}{s_i}\cdot(\Delta R_i)^2,\\
\alpha^* = \underset{\alpha \in (0,1)}{\arg\min} ~\mathcal{L}_\text{pref}+\lambda\cdot\mathcal{L}_\text{tie},\label{equation:eq13}
\end{gather}
where $\lambda$ controls the strength of the indifference constraint. This formulation enables robust calibration of $\alpha$ to align the proposed RAS with human reliability preferences.

\section{Training Abstention-Aware ASR}

Our training pipeline is designed in two consecutive stages to enhance the model's uncertainty awareness and self-correction capabilities.

\subsection{Stage 1: placeholder supervision (PH-Supv)}\label{sec:ph-supv}

The primary objective of this stage is to construct a dataset that guides the base model to identify and flag prediction errors.
{\fontfamily{ppl}\selectfont
\begin{figure}[t]
  \centering
  \tcbox[
    colback=blue!15!white,        
    colframe=black,         
    rounded corners,        
    boxrule=0.8pt,          
    arc=3mm,                
    left=1pt,right=1pt,top=1pt,bottom=1pt  
  ]{%
    \resizebox{0.94\columnwidth}{!}{%
      {\bfseries
      \begin{tabular}{l|c|c|c|c|c|c|c}
        $\bm{y_{gt}}$: & \textcolor{red}{} & \text{chronic disease of} & \textcolor{red}{hair} & \textcolor{red}{follicles} & \text{and} & \textcolor{red}{sebaceous} & \text{gland} \\
        $\bm{y_{hyp}}$: & \textcolor{red}{the} & \text{chronic disease of} & \textcolor{red}{her} &  & \text{and} & \textcolor{red}{spoculus} & \text{gland} \\
        \text{type:}  & Insert & Match & Substitute & Delete & Match & Substitute & Match \\
        $\bm{y_{ph}}$:  & \texttt{<ph>} & chronic disease of & \texttt{<ph>} & \texttt{<ph><ph><ph>} & and & \texttt{<ph><ph><ph>} & gland \\
      \end{tabular}
      }
    }%
  }\vspace{-10pt}
  \caption{Strategy for generating Stage 1 $y_{ph}$. Incorrectly predicted segments are replaced with $\mathcal{PH}$.\vspace{-5pt}}
  \label{fig:replace}
\end{figure}
}
\subsubsection{Training set construction}\label{trainset_con}

Given an audio input $a$ and its ground truth transcription $y_{gt}$, we first conduct inference using the base model $\mathcal{M}$ to generate a hypothesis $y_{hyp}$. To locate recognition errors, we align $y_{hyp}$ with $y_{gt}$ using the standard Word Error Rate (WER) calculation.


This alignment yields a sequence of operations, where each operation is one of $\{\text{Match, Substitute, Insert, Delete}\}$. We define a GT-guided $\mathcal{PH}$-replacement sequence $y_{ph}$, constructed from these operations as follows:\\
\textbf{Match:} The corresponding text in $y_{hyp}$ is retained in $y_{ph}$.\\
\textbf{Substitute / Insert:} The erroneous text segment $t_{err}$ in $y_{hyp}$ is replaced by the special token $\mathcal{PH}$. To maintain relative sequence length, the number of $\mathcal{PH}$ tokens is determined by the tokenizer $\mathcal{T}$ of the base model. Specifically, we insert $N$ tokens of $\mathcal{PH}$, where $N = |\mathcal{T}(t_{err})|$.\\
\textbf{Delete:} In this case, $y_{hyp}$ omits text present in $y_{gt}$. We insert $\mathcal{PH}$ tokens into $y_{ph}$ corresponding to the missing text segment $t_{miss}$ from $y_{gt}$. Similar to the above, the number of $\mathcal{PH}$ tokens is set to $N = |\mathcal{T}(t_{miss})|$.

Figure \ref{fig:replace} shows an example of the process above. This process results in a refined training set consisting of pairs $(a, y_{ph})$.

\subsubsection{Model training}
Prior to training, we expand the vocabulary of the base model $\mathcal{M}$ to include the new token $\mathcal{PH}$, ensuring that the tokenizer treats it as a single, indivisible token. The model is then fine-tuned on the constructed dataset using the same ASR objective as $\mathcal M$.
In this work, we employ Whisper \cite{whisper} as the base model $\mathcal M$, and thus use a cross entropy objective in this stage.
\subsection{Stage 2: group relative policy optimization (RL)}
\label{sec:grpo}
Once the model acquires the capability to output $\mathcal{PH}$, we employ Group Relative Policy Optimization (GRPO) \cite{grpo} algorithm to optimize its output policy,  using utterance-level RAS proposed in Section \ref{sec:utility} as the reward signal.
Specifically, for each input prompt $q$, GRPO samples a group of $G$ outputs $\{o_1, o_2, \ldots, o_G\}$ from the current policy $\pi_{\theta_{\text{old}}}$. Each output $o_i$ is evaluated by the RAS reward function to obtain a reward $r_i$. The group-relative advantage is then computed as: $\hat{A}_i = \frac{r_i - \mathrm{mean}(\{r_j\}_{j=1}^{G})}{\mathrm{std}(\{r_j\}_{j=1}^{G})}.$
The policy is optimized by maximizing the following objective:
\begin{equation}
    \mathcal{L}_{\text{GRPO}}(\theta) = \mathbb{E}\left[ \frac{1}{G} \sum_{i=1}^{G} \frac{1}{|o_i|} \sum_{t=1}^{|o_i|} \left( \hat{L}_{i,t} - \beta\, D_{\mathrm{KL}}^{i,t} \right) \right],
\end{equation}
\begin{equation}    
    \hat{L}_{i,t} =  \min\left\{\rho_{i,t}\hat{A}_i,~ \mathrm{clip}(\rho_{i,t}, 1{-}\epsilon, 1{+}\epsilon) \hat{A}_i\right\},
\end{equation}
where $\rho_{i,t} = \frac{\pi_\theta(o_{i,t} \mid q,\, o_{i,<t})}{\pi_{\theta_{\text{old}}}(o_{i,t} \mid q,\, o_{i,<t})}$ is the importance sampling ratio, $\epsilon$ is the clipping parameter, and $D_{\mathrm{KL}}^{i,t}$ is the per-token KL divergence against the reference policy $\pi_{\text{ref}}$ as defined in \cite{grpo}.





\section{Experiments}
\label{sec:experiments}

\subsection{Datasets}
We conduct experiments on two datasets: LibriSpeech \cite{librispeech}, a widely used English audiobook corpus, and the TALCS Corpus \cite{talcs}, an English-Mandarin code-switching dataset. For LibriSpeech, we utilize train-clean-360 for training and test-clean for evaluation. To assess ASR reliability under adverse acoustic conditions, we simulate a noisy variant (Noisy LibriSpeech) by injecting Additive White Gaussian Noise on the total duration of original audio samples. Four distinct train and test subsets were generated with Signal-to-Noise Ratios (SNRs) fixed at $\{0, 5, 10, 20\}$ dB, respectively. For all other datasets, we adhere to the official training, development, and test partitions.

\subsection{Evaluation metric and human alignment results}

We adopt RAS (defined in Section \ref{sec:utility}) as our evaluation criterion, with $\alpha$ estimated by human alignment listening test.
The test design follows Section \ref{section:listening-test}. Audio samples were manually selected from the Medical ASR Recording Dataset \cite{medical} (173 samples) and the AMI Corpus \cite{ami-corpus} (191 samples), covering terminology-intensive medical speech and noise-corrupted conversational speech. For each sample, two transcript variants were constructed as described.

After validity filtering, 980 preference annotations were collected from 42 participants. The human oracle upper bound, computed via majority voting and annotator agreement with the consensus, reached 78.11\%, indicating strong inter-annotator consistency. With $\lambda=0.1$ (corresponding to a low tie rate of 6.63\%), minimizing the objective in Equation \eqref{equation:eq13} yields $\alpha = 0.5064$ and an average RAS gap of $\Delta U = 0.0461$, demonstrating alignment between the proposed RAS formulation and human reliability judgments.
\subsection{Experimental setup}
We use the following experiment notations:\\
\textbf{Base \& Base+Logit}: We use Whisper-Tiny as the baseline (Base). Base+Logit, which can be considered as a confidence baseline, applies logit-based $\mathcal{PH}$-replacement using token-level confidence aggregated multiplicatively (word-level for English; character/word-level for Chinese), following official Whisper heuristics\footnote{\url{https://github.com/openai/whisper/discussions/1183}}. For each dataset, we replace tokens below a confidence \textit{bar} with $\mathcal{PH}$, tuning \textit{bar} (typically in $[0.1, 0.3]$) to maximize RAS.\\
\textbf{Base+PH-Supv}: This stage uses $\mathcal{PH}$ replacements from Base predictions on the training set (details in Section \ref{sec:ph-supv}). We fine-tune the decoder and embeddings for 8 epochs using AdamW~\cite{adamw} with a batch size of 64, a $1.0\times10^{-5}$ learning rate, 1,000 warmup steps, and linear decay.\\
\textbf{Base+PH-Supv+RL}: Initialized from Base+PH-Supv, we further optimize $\mathcal{PH}$ supervision via GRPO (Section \ref{sec:grpo}). We unfreeze the decoder/embeddings and train with a global batch size of 64. For each prompt, $G=8$ responses are sampled (512 samples/step) using temperature 0.7 and top-$p$ 0.95. We employ an adaptive KL penalty \cite{schulman2017proximalpolicyoptimizationalgorithms} with $\beta_0=0.2$, updated every 50 steps: $\beta_{t+1} = \beta_t \exp(\eta (\textit{KL}_t - \textit{KL}_\text{target}))$, where $\eta=0.02$ and $\textit{KL}_\text{target}=30$. We use the Adam optimizer \cite{adam} with a $2\times10^{-6}$ peak learning rate (1,000-step warmup and linear decay) and early stopping based on the stabilization of the reward mean.\\
\textbf{GT-guided $\mathcal{PH}$-replacement}: We construct $\mathcal{PH}$-replaced sequences by applying the algorithm in Section~\ref{trainset_con} to Base outputs, guided by test set ground-truth.
\subsection{Main results}\label{sec:results}
\begin{table}[H]
  \centering
  \setlength{\tabcolsep}{2.5pt}
  \caption{RAS performance on LibriSpeech and TALCS. \textbf{Boldface} marks the best result per dataset, excluding GT-guided $\mathcal{PH}$-replacement systems.\vspace{-1pt}}
  
  \resizebox{\columnwidth}{!}{%
  
  \begin{tabular}{l ccc ccc}
    \toprule
    \multicolumn{1}{c}{\multirow{2.5}{*}{\textbf{Method}}} & \multicolumn{3}{c}{\textbf{LibriSpeech}} & \multicolumn{3}{c}{\textbf{TALCS}} \\
    \cmidrule(lr){2-4}\cmidrule(lr){5-7}
    & \textit{RAS}$\uparrow$ & \textit{Usefulness}$\uparrow$ & \textit{Cost}$\downarrow$ & \textit{RAS}$\uparrow$ & \textit{Usefulness}$\uparrow$ & \textit{Cost}$\downarrow$ \\
    \midrule
    \textit{Base}                    & 0.8603          & 0.9362 & 0.0759 & $-$0.1093       & 0.5874 & 0.6968 \\
    \textit{Base+Logit}              & 0.8650          & 0.9349 & 0.0698 & $-$0.0650       & 0.5595 & 0.6245 \\
    \textit{Base+PH-Supv+RL (Ours)}  & \textbf{0.8811} & \textbf{0.9376}     & \textbf{0.0565}     & \textbf{0.4786} & \textbf{0.7391}     & \textbf{0.2940}     \\
    \midrule
    \textit{GT-guided $\mathcal{PH}$-replacement}                  & 0.9031          & 0.9361 & 0.0329 & 0.3772          & 0.5874 & 0.2103 \\
    \bottomrule
  \end{tabular}}
  \label{tab:noise_clean}
\end{table}

\begin{figure}[!t]
    \centering
    \includegraphics[width=0.8\columnwidth]{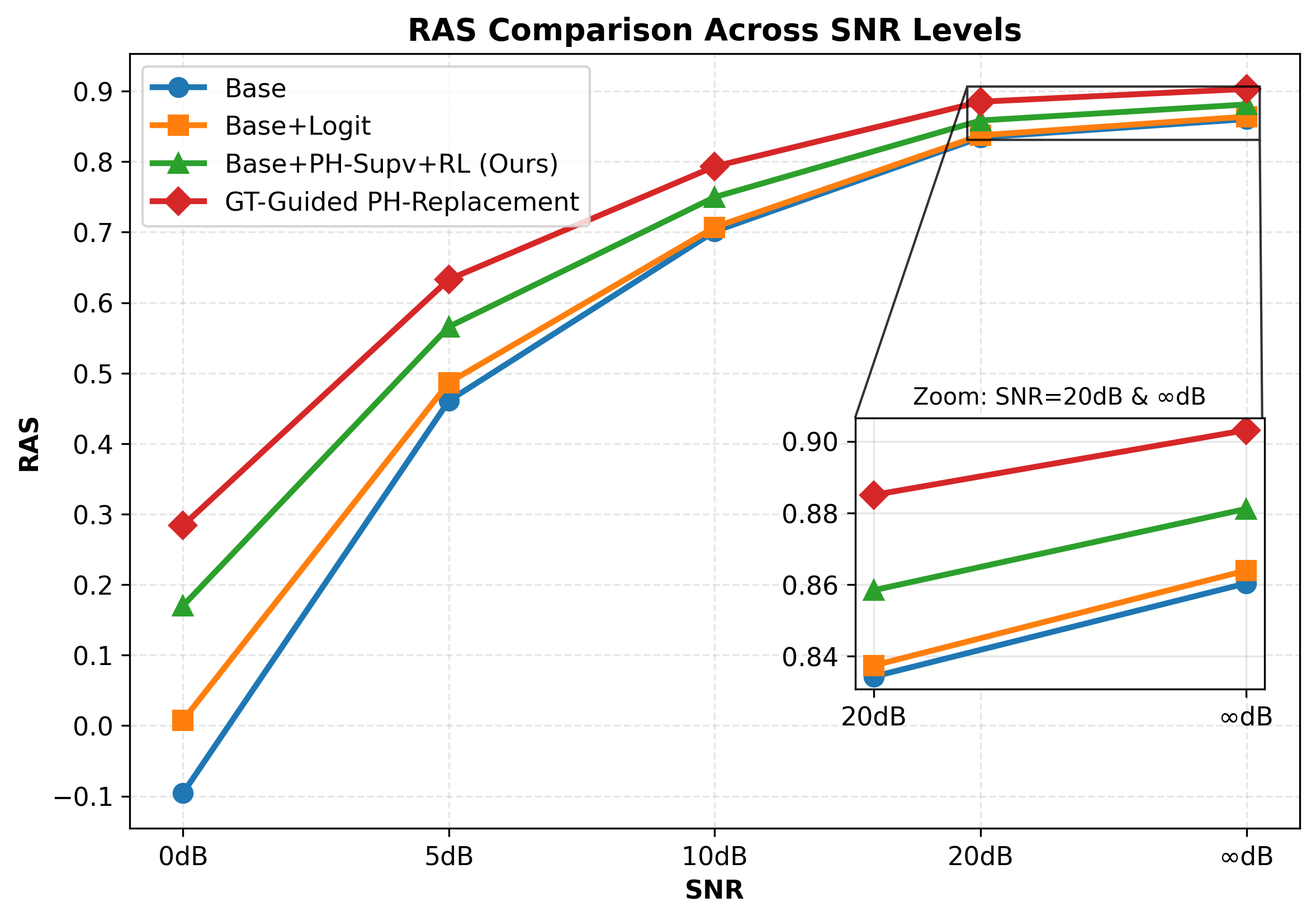}\vspace{-10pt}
    \caption{RAS on Noisy LibriSpeech. \vspace{-7pt}
    }
    \label{fig:noise}
\end{figure}

Table~\ref{tab:noise_clean} reports results under clean (LibriSpeech) and code-switching (TALCS) conditions. Our method \textit{Base+PH-Supv+RL} consistently outperforms both baselines across all metrics. On TALCS, where the base model yields a negative RAS ($-0.11$), our method achieves $0.48$, demonstrating that $\mathcal{PH}$-supervised pretraining followed by RL is especially effective in challenging code-switching scenarios. The confidence baseline (\textit{Base+Logit}) provides marginal gains, confirming that naive confidence thresholding is insufficient. \textit{GT-guided $\mathcal{PH}$-replacement} serves as a near-oracle upper bound guided by ground-truth.

Figure~\ref{fig:noise} reports the performance under noisy conditions (Noisy LibriSpeech). At $\text{SNR}=0$ dB, our method improves RAS by $0.2657$ over \textit{Base}, compared to $0.0208$ improvement at clean conditions, highlighting the robustness benefit in highly noisy environments.
The improvement in RAS brought by our method becomes increasingly pronounced as the SNR decreases, demonstrating that the proposed training pipeline, i.e., PH-Supv+RL, confers greater robustness gains in more challenging acoustic environments.

\subsection{Ablation study}
\begin{table}[H]
  \centering
    \setlength{\tabcolsep}{2.5pt}
  \caption{Ablation study for the two-stage training pipeline.}
  \resizebox{\columnwidth}{!}{%
  \begin{tabular}{l ccc ccc}
    \toprule
    \multicolumn{1}{c}{\multirow{2.5}{*}{\textbf{Method}}} & \multicolumn{3}{c}{\textbf{LibriSpeech}} & \multicolumn{3}{c}{\textbf{TALCS}} \\
    \cmidrule(lr){2-4}\cmidrule(lr){5-7}
    \cmidrule(lr){2-4}\cmidrule(lr){5-7}
    & \textit{RAS}$\uparrow$ & \textit{Usefulness}$\uparrow$ & \textit{Cost}$\downarrow$ & \textit{RAS}$\uparrow$ & \textit{Usefulness}$\uparrow$ & \textit{Cost}$\downarrow$ \\
    \midrule
    \textit{Base+PH-Supv}       & 0.8696          & 0.9277 & 0.0581 & 0.4054          & 0.6520 & \textbf{0.2466} \\
    \textit{Base+PH-Supv+RL}    & \textbf{0.8811} & \textbf{0.9376}     & \textbf{0.0565}     & \textbf{0.4786} & \textbf{0.7391}     & 0.2940     \\
    \bottomrule
  \end{tabular}}
  \label{tab:grpo_clean}
\end{table}
Table~\ref{tab:grpo_clean} presents an ablation study isolating the contribution of GRPO-based RL training.
Comparing \textit{Base+PH-Supv} and \textit{Base+PH-Supv+RL}, we observe that RL consistently
improves RAS and Usefulness on both datasets. On TALCS, while Cost slightly increases, this reflects the RL objective's optimization toward RAS: the model
learns that the Usefulness gain from predicting more words outweighs the Cost incurred when
some $\mathcal{PH}$-replaced tokens are instead decoded incorrectly. The net effect is a substantially
higher RAS, confirming that the GRPO-based RL stage provides
complementary benefits on top of supervised placeholder training.

Notably, on TALCS both \textit{Base+PH-Supv} and \textit{Base+PH-Supv+RL} in Table \ref{tab:grpo_clean} surpass the GT-guided $\mathcal{PH}$-replacement upper bound in Table \ref{tab:noise_clean}.
Since Base performs poorly there, GT-guided replacement is fundamentally limited by its weak code-switching capability.
\textit{PH-Supv} training, by contrast, genuinely improves the model: even without full GT transcripts, the correct tokens retained after replacing Base's errors provide sufficient supervision to substantially boost TALCS performance, with RL further amplifying this advantage.

\section{Conclusion}
This work redefines ASR reliability by introducing placeholder-based abstention, shifting the paradigm from speculative transcription to risk-aware reporting. We propose the RAS, a principled metric calibrated via human preferences to balance informativeness with error aversion. By implementing an abstention-aware training pipeline, we demonstrate that models can substantially improve reliability without sacrificing competitive accuracy. Our framework establishes a new criterion for trustworthy speech processing, from which both downstream applications and human users benefit. We expect the RAS framework to provide a robust evaluation foundation for future research in trustworthy speech processing, particularly in serving as a reliable reward signal for optimizing model behavior through reinforcement learning.


\section{Acknowledgments}
This work has been supported by the China NSFC Project (No. 92370206).





\section{Generative AI Use Disclosure}
Generative AIs are used in this work for manuscript polishing and code troubleshooting. We utilize these tools to improve linguistic clarity and assist in technical debugging. The conceptual framework, experimental design, and final writing are entirely conducted by human authors, who take full responsibility for the content and consent to its submission.




\bibliographystyle{IEEEtran}
\bibliography{main}

\begin{thebibliography}{10}
\providecommand{\url}[1]{#1}
\csname url@samestyle\endcsname
\providecommand{\newblock}{\relax}
\providecommand{\bibinfo}[2]{#2}
\providecommand{\BIBentrySTDinterwordspacing}{\spaceskip=0pt\relax}
\providecommand{\BIBentryALTinterwordstretchfactor}{4}
\providecommand{\BIBentryALTinterwordspacing}{\spaceskip=\fontdimen2\font plus
\BIBentryALTinterwordstretchfactor\fontdimen3\font minus \fontdimen4\font\relax}
\providecommand{\BIBforeignlanguage}[2]{{%
\expandafter\ifx\csname l@#1\endcsname\relax
\typeout{** WARNING: IEEEtran.bst: No hyphenation pattern has been}%
\typeout{** loaded for the language `#1'. Using the pattern for}%
\typeout{** the default language instead.}%
\else
\language=\csname l@#1\endcsname
\fi
#2}}
\providecommand{\BIBdecl}{\relax}
\BIBdecl

\bibitem{Optimal-Reject}
C.~K. Chow, ``An optimum character recognition system using decision functions,'' \emph{IRE Transactions on Electronic Computers}, vol. EC-6, no.~4, pp. 247--254, 1957.

\bibitem{Calibrated-Structured-Prediction}
V.~Kuleshov and P.~Liang, ``Calibrated structured prediction,'' in \emph{Proceedings of the 29th International Conference on Neural Information Processing Systems - Volume 2}, ser. NIPS'15.\hskip 1em plus 0.5em minus 0.4em\relax Cambridge, MA, USA: MIT Press, 2015, p. 3474–3482.

\bibitem{xu2024reducing}
H.~Xu, Z.~Zhu, L.~Pan, Z.~Wang, S.~Zhu, D.~Ma, R.~Cao, L.~Chen, and K.~Yu, ``Reducing tool hallucination via reliability alignment,'' in \emph{Proceedings of the 42nd International Conference on Machine Learning}, ser. Proceedings of Machine Learning Research, vol. 267.\hskip 1em plus 0.5em minus 0.4em\relax PMLR, 13--19 Jul 2025, pp. 69\,992--70\,006.

\bibitem{xu2024rejection}
H.~Xu, Z.~Zhu, S.~Zhang, D.~Ma, S.~Fan, L.~Chen, and K.~Yu, ``Rejection improves reliability: Training {LLM}s to refuse unknown questions using {RL} from knowledge feedback,'' in \emph{First Conference on Language Modeling}, 2024.

\bibitem{Schnwlder2025AbstentionIA}
E.~Sch{\"o}nw{\"a}lder, C.~Falkenberg, C.~Hartmann, and W.~Lehner, ``Abstention is all you need,'' \emph{2025 IEEE 12th International Conference on Data Science and Advanced Analytics (DSAA)}, pp. 1--10, 2025.

\bibitem{zheng2025enhancing}
H.~Zheng, H.~Xu, Y.~Liu, S.~Fan, L.~Chen, P.~Fung, and K.~Yu, ``Enhancing {LLM} reliability via explicit knowledge boundary modeling,'' in \emph{Second Conference on Language Modeling}, 2025.

\bibitem{Jiang2005ConfidenceMF}
H.~Jiang, ``Confidence measures for speech recognition: A survey,'' \emph{Speech Communications}, vol.~45, pp. 455--470, 2005.

\bibitem{Onea2021AnEO}
D.~Oneaţă, A.~Caranica, A.~Stan, and H.~Cucu, ``An evaluation of word-level confidence estimation for end-to-end automatic speech recognition,'' \emph{2021 IEEE Spoken Language Technology Workshop (SLT)}, pp. 258--265, 2021.

\bibitem{Futami2021ASRRA}
H.~Futami, H.~Inaguma, M.~Mimura, S.~Sakai, and T.~Kawahara, ``{ASR} rescoring and confidence estimation with electra,'' \emph{2021 IEEE Automatic Speech Recognition and Understanding Workshop (ASRU)}, pp. 380--387, 2021.

\bibitem{Naowarat2023WordlevelCE}
B.~Naowarat, T.~Kongthaworn, and E.~Chuangsuwanich, ``{Word-level Confidence Estimation for {CTC} Models},'' in \emph{{Interspeech 2023}}, 2023, pp. 3297--3301.

\bibitem{Huo2025IdentifyingAC}
M.~Huo, Y.~Zhang, and Y.~Tang, ``Identifying and calibrating overconfidence in noisy speech recognition,'' \emph{arXiv preprint arXiv:2509.07195}, 2025.

\bibitem{lawrence-overview}
L.~R. Rabiner, \emph{A tutorial on hidden Markov models and selected applications in speech recognition}.\hskip 1em plus 0.5em minus 0.4em\relax San Francisco, CA, USA: Morgan Kaufmann Publishers Inc., 1990, p. 267–296.

\bibitem{nist_hub5ne_1997}
{National Institute of Standards and Technology}, ``The 1997 hub-5ne evaluation plan for recognition of conversational speech over the telephone,'' \url{https://catalog.ldc.upenn.edu/docs/LDC2002S25/hub5nev3.htm}, 1997, accessed Feb 14, 2026.

\bibitem{edit-distance}
V.~I. Levenshtein, ``Binary codes capable of correcting deletions, insertions, and reversals,'' \emph{Soviet physics. Doklady}, vol.~10, pp. 707--710, 1965.

\bibitem{K2024AdvocatingCER}
D.~Thennal, J.~James, D.~P. Gopinath \emph{et~al.}, ``Advocating character error rate for multilingual {ASR} evaluation,'' in \emph{Findings of the Association for Computational Linguistics: NAACL 2025}, 2025, pp. 4926--4935.

\bibitem{MER_WIL}
A.~C. Morris, V.~Maier, and P.~Green, ``{From WER and RIL to MER and WIL: improved evaluation measures for connected speech recognition},'' in \emph{{Interspeech 2004}}, 2004, pp. 2765--2768.

\bibitem{bleu}
K.~Papineni, S.~Roukos, T.~Ward, and W.-J. Zhu, ``{BLEU}: a method for automatic evaluation of machine translation,'' in \emph{Proceedings of the 40th Annual Meeting of the Association for Computational Linguistics}.\hskip 1em plus 0.5em minus 0.4em\relax Philadelphia, Pennsylvania, USA: Association for Computational Linguistics, Jul. 2002, pp. 311--318.

\bibitem{ROUGE}
C.-Y. Lin, ``{ROUGE}: A package for automatic evaluation of summaries,'' in \emph{Text Summarization Branches Out}.\hskip 1em plus 0.5em minus 0.4em\relax Barcelona, Spain: Association for Computational Linguistics, Jul. 2004, pp. 74--81.

\bibitem{bertscore}
T.~Zhang, V.~Kishore, F.~Wu, K.~Q. Weinberger, and Y.~Artzi, ``{BERTScore}: Evaluating text generation with {BERT},'' in \emph{International Conference on Learning Representations}, 2020.

\bibitem{whisper}
A.~Radford, J.~W. Kim, T.~Xu, G.~Brockman, C.~Mcleavey, and I.~Sutskever, ``Robust speech recognition via large-scale weak supervision,'' in \emph{Proceedings of the 40th International Conference on Machine Learning}, ser. Proceedings of Machine Learning Research, vol. 202.\hskip 1em plus 0.5em minus 0.4em\relax PMLR, 23--29 Jul 2023, pp. 28\,492--28\,518.

\bibitem{sutton1998reinforcement}
R.~S. Sutton, A.~G. Barto \emph{et~al.}, \emph{Reinforcement learning: An introduction}.\hskip 1em plus 0.5em minus 0.4em\relax MIT press Cambridge, 1998, vol.~1, no.~1.

\bibitem{librispeech}
V.~Panayotov, G.~Chen, D.~Povey, and S.~Khudanpur, ``Librispeech: An asr corpus based on public domain audio books,'' in \emph{2015 IEEE International Conference on Acoustics, Speech and Signal Processing (ICASSP)}, 2015, pp. 5206--5210.

\bibitem{xu-etal-2025-alignment}
H.~Xu, Z.~Wang, Z.~Zhu, L.~Pan, X.~Chen, S.~Fan, L.~Chen, and K.~Yu, ``Alignment for efficient tool calling of large language models,'' in \emph{Proceedings of the 2025 Conference on Empirical Methods in Natural Language Processing}, C.~Christodoulopoulos, T.~Chakraborty, C.~Rose, and V.~Peng, Eds.\hskip 1em plus 0.5em minus 0.4em\relax Suzhou, China: Association for Computational Linguistics, Nov. 2025, pp. 17\,776--17\,792.

\bibitem{beaqlejs}
S.~Kraft and U.~Zölzer, ``{BeaqleJS}: {HTML5} and {JavaScript} based framework for the subjective evaluation of audio quality,'' in \emph{Proceedings of the Linux Audio Conference}, Karlsruhe, Germany, May 2014.

\bibitem{bradley-terry}
R.~A. Bradley and M.~E. Terry, ``Rank analysis of incomplete block designs: I. the method of paired comparisons,'' \emph{Biometrika}, vol.~39, no. 3/4, pp. 324--345, 1952.

\bibitem{grpo}
Z.~Shao, P.~Wang, Q.~Zhu, R.~Xu, J.~Song, X.~Bi, H.~Zhang, M.~Zhang, Y.~Li, Y.~Wu \emph{et~al.}, ``Deepseekmath: Pushing the limits of mathematical reasoning in open language models,'' \emph{arXiv preprint arXiv:2402.03300}, 2024.

\bibitem{talcs}
C.~Li, S.~Deng, Y.~Wang, G.~Wang, Y.~Gong, C.~Chen, and J.~Bai, ``{TALCS: An open-source Mandarin-English code-switching corpus and a speech recognition baseline},'' in \emph{{Interspeech 2022}}, 2022, pp. 1741--1745.

\bibitem{medical}
\BIBentryALTinterwordspacing
{Hani89}, ``Medical {ASR} recording dataset,'' Hugging Face Datasets, 2023. [Online]. Available: \url{https://huggingface.co/datasets/Hani89/medical_asr_recording_dataset}
\BIBentrySTDinterwordspacing

\bibitem{ami-corpus}
I.~Mccowan, J.~Carletta, W.~Kraaij, S.~Ashby, S.~Bourban, M.~Flynn, M.~Guillemot, T.~Hain, J.~Kadlec, V.~Karaiskos, M.~Kronenthal, G.~Lathoud, M.~Lincoln, A.~Lisowska~Masson, W.~Post, D.~Reidsma, and P.~Wellner, ``The {AMI} meeting corpus,'' \emph{Int'l. Conf. on Methods and Techniques in Behavioral Research}, 01 2005.

\bibitem{adamw}
I.~Loshchilov and F.~Hutter, ``Decoupled weight decay regularization,'' in \emph{International Conference on Learning Representations}, 2019.

\bibitem{schulman2017proximalpolicyoptimizationalgorithms}
J.~Schulman, F.~Wolski, P.~Dhariwal, A.~Radford, and O.~Klimov, ``Proximal policy optimization algorithms,'' \emph{arXiv preprint arXiv:1707.06347}, 2017.

\bibitem{adam}
D.~P. Kingma and J.~Ba, ``Adam: A method for stochastic optimization,'' in \emph{International Conference on Learning Representations (ICLR)}, 2015.

\end{thebibliography}

\end{document}